# Possible hydrogen doping and enhancement of $T_c$ (= 35 K) in a LaFeAsO-based superconductor

K. Miyazawa[1,2], S. Ishida[1,3], K. Kihou[1,5], P. M. Shirage[1], M. Nakajima[1,3], C. H. Lee[1,5], H. Kito[1,5], Y. Tomioka[1,5], T. Ito[1,5], H. Eisaki[1,5], H. Yamashita[4], H. Mukuda[4,5], K. Tokiwa[2], S. Uchida[3,5] and A. Iyo[1, 2, 5]

[1]National Institute of Advanced Industrial Science and Technology, Tsukuba, Ibaraki 305-8568, Japan

[2]Department of Applied Electronics, Tokyo University of Science, Noda, Chiba 275-8510, Japan

[3]Department of Physics, University of Tokyo, Tokyo 113-0033, Japan

[4]Graduate School of Engineering Science, Osaka University, Toyonaka, Osaka 560-8531, Japan

[5]JST, Transformative Research-Project on Iron Pnictides (TRIP), 5, Sanbancho, Chiyoda, Tokyo 102-0075, Japan

**Abstract:**

We report that the incorporation of hydroxide ions $(OH)^-$ significantly enhances the superconducting transition temperature ($T_c$) in the *Ln*FeAsO-based superconductors (*Ln*1111: *Ln* = La, Ce and Pr). For La1111, $T_c$ of the $(OH)^-$ incorporated sample synthesized using high-pressure technique becomes 35 K, which is higher by 7 K than the typical optimally-doped La1111 superconductors. Similar enhancement in $T_c$ is also observed for Ce1111 and Pr1111. $^1$H-NMR measurement have confirmed the existence of hydrogen atoms in the samples. Accompanying the $(OH)^-$ incorporation, the lattice parameters are largely contracted, down to the values which have never been attained by any other dopings/substitutions.

Discovery of the superconductivity at 26 K in LaFeAs(O,F) has demonstrated that there still exists a vast frontier to quest for high-transition temperature (high-$T_c$) superconductors [1]. To date, various kinds of Fe(Ni)-based superconductors have been synthesized and the maximum $T_c$ has exceeded 50 K [2]. Superconductivity in the Fe-based compounds usually shows up by the chemical substitution on their parent compounds. For the *Ln*FeAsO-based superconductors (*Ln*1111: *Ln*=rare earth), partial replacement of O by F [2], partial replacement of *Ln* by Th [3], or introduction of O-deficiency [4,5] yields highest $T_c$. Superconductivity is also realized through the substitution within the FeAs layers, such as Co at the Fe site [6] or P at the As site [7], while their $T_c$'s are comparatively lower.

Considering the rich variety in the chemical substitution, one would naively expect that superconductors, possibly with higher $T_c$'s, can be synthesized by developing substitution methods, even starting from the known parent compounds. Based on this motivation, we have tried to dope hydrogen (H) ions into the O-deficient *Ln*FeAsO (*Ln*1111: *Ln*=La, Ce, Pr) samples by incorporating rare-earth hydroxides $Ln(OH)_3$ and have found that $T_c$ of La1111 increases from 28 K to 35 K.

Polycrystalline samples are synthesized using a cubic-anvil-type high-pressure (HP) apparatus [8]. Starting materials are powders of Fe, $Fe_2O_3$, As, *Ln*As (*Ln* = La, Ce and Pr), and $La(OH)_3$. The following two methods were employed. In the first method (method-I), all the starting materials are mixed in the formula of $LaFeAsO_{1-y}H_x$ ($x$ ~ 0.6 and $y$ ~ 0.4). The mixed powder was ground and pressed into a pellet, loaded in a BN crucible surrounded by a graphite heater sleeve and a pyrophyllite cube used as a HP medium (Fig. 1 (a)). In the second method (method-II), multiple sample pellets are simultaneously loaded within a BN crucible (Fig. 1(b)). A H-free target sample with its nominal composition $LnFeAsO_{1-y}$ (*Ln* = La, Ce and Pr, $y$ ~ 0.4) is placed in the middle, and two other samples which include H in the formula of $LaFeAsO_{0.6}H_{0.6}$ are located at the outer sides, separated from the central one by BN pellets. In method-II, we expect that the transfer of H from the outer pellets to the inner one takes place during the synthesis. As shall be seen later, the two methods indeed yield the same results. The pellets are heated at 1100 °C under a pressure of 3.5 GPa for 2 h. The samples thus synthesized are abbreviated as *Ln*1111(H) hereafter.

Powder X-ray diffraction (XRD) patterns were measured using $CuK_{\alpha}$ radiation. Proton nuclear magnetic resonance ($^{1}$H-NMR) measurements were carried out in order to confirm the presence of H atoms in the samples. The dc magnetic susceptibility was measured using a SQUID magnetometer (Quantum Design MPMS) under a magnetic field of 5 Oe. The resistivity was measured by a four-probe method (Quantum Design PPMS).

XRD patterns for the samples with nominal compositions of $LaFeAsO_{0.6}H_{0.6}$ (method-I), $LaFeAsO_{0.55}$, $CeFeAsO_{0.65}$ and $PrFeAsO_{0.63}$ (method-II), are shown in Fig. 1(c). All the peaks can be indexed based on the *Ln*FeAsO crystal structure. As shown later, *a*- and *c*-axis lattice parameters are dramatically contracted compared to the H-free *Ln*1111.

$^{1}$H-NMR spectrum for La1111(H) synthesized by method-I is shown in Fig. 2(a). The spectrum is composed of a symmetric and single peak, suggesting that the precipitation of stable impurity phase including H atoms is negligible. The half-width of the spectrum is approximately 90 Oe, much broader than in the compounds including H on the stoichiometry. In Fig. 2(b) we show the $^{1}$H-NMR spectrum for $Ag_6O_8Ag(HF_2)$, as such an example [9]: The typical half-width is as narrow as 2.6 Oe in this compound, since H atoms are located at a regular crystallographic site. Thus, the broad $^{1}$H-NMR spectral width of La1111(H) is attributed to the wide distribution of transferred hyperfine fields from Fe atoms in association with some local disorder created by the H atoms themselves.

The temperature ($T$) dependence of the zero-field-cooling (ZFC) and the field-cooling (FC) magnetic susceptibility ($\chi$) for the La1111(H) samples synthesized using method-I and II are shown in Fig. 3(a) and (b), respectively, together with that for the oxygen deficient sample for comparison in Fig. 3(a). Both data show the sharp superconducting transition at 35 K (see inset), which is much higher than those of the H-free, F-substituted or O-deficient La1111 superconductors ($T_c$ = 28 K). The superconducting volume fraction estimated from the ZFC magnetic susceptibility at 5 K is 78 % (method-I) and 56 % (method-II), respectively, ensuring their bulk superconductivity. Similarly, $T_c$'s for Ce1111(H) (Fig. 3(c)) and Pr1111(H) (Fig. 3(d)) increases up to 46 K and 50 K, which are higher than those for H-free *Ln*1111 by 7 K for *Ln* = Ce and by 2 K for Pr, respectively.

Fig. 4 show the *T*-dependent resistivity ($\rho$) for *Ln*1111(H) (*Ln* = La, Ce and Pr) samples synthesized using method-II. All the samples show the metallic behavior ($d\rho/dT > 0$) and exhibit sharp superconducting transitions. The onset temperature of the transition are about 39, 48 and 52 K ($T_c$ ($\rho$-onset)), and the zero-resistivity is observed at about 34, 45 and 49 K ($T_c$ ($\rho$-zero)), for *Ln* = La, Ce and Pr, respectively. Note that $T_c$ ($\rho$-zero) and $T_c$ ($\rho$-onset) are very close to $T_c$ ($\chi$-onset), indicating the good homogeneity of our samples. The $T_c$ values are listed in Table I with the lattice parameters.

In Fig. 5, the *a*- and *c*- lattice parameters of the *Ln*1111(H) samples calculated from the XRD patterns and $T_c$ ($\chi$-onset) are plotted, together with those of H-free, O-deficient *Ln*1111 [10] and their parent compounds. The lattice parameters of *Ln*1111(H) become substantially shorter compared to the H-free samples. In particular, the *a*-axis lattice parameter of La1111(H) becomes even shorter than that of the optimally-doped $CeFeAsO_{1-y}$ ($T_c$ = 38 K). Now let us consider the reason for the enhancement in $T_c$. In *Ln*1111, there is an empirical correlation between $T_c$ and their lattice parameters. In the inset of Fig. 5, we show the relationship between $T_c$ ($\chi$-onset) and the *a*-axis lattice parameters of *Ln*1111(H), together with O-deficient *Ln*1111 (*Ln* = La, Ce, Pr, Nd, Sm, Gd, Tb and Dy) [10]. The results on *Ln*1111(H) are quantitatively fitted into the general trend established in the H-free, O-deficient *Ln*1111. The agreement between the two cases strongly suggests that the enhancement of $T_c$ in *Ln*1111(H) is effectively caused by the lattice contraction caused by the H incorporation.

Next, let us consider the reason why the incorporation of H (or OH) results in contracting the lattice parameters. Recently, water-induced superconductivity was reported by Hiramatsu *et al.* on an undoped $SrFe_2As_2$ thin film [11]. In their samples they have also observed the shrinkage of *c*-axis lattice parameter. They assume that either $H_2O$, OH, or O molecules/ions are intercalated at the interstitial sites. Such intercalation is less likely in *Ln*1111(H), since the intercalation of large molecules/atoms should results in expanding the lattice volume, which is opposite to the observation.

It is more likely that the $H^+$ ions are intercalated. The ionic radius of $H^+$ is estimated as -0.38 Å [12], which means that the lattice parameters are contracted by the incorporation of $H^+$ ions. It would be possible that the $H^+$ ions are inserted into the interstitial site and attract the neighboring $O^{2-}$ and/or $As^{3-}$ ions, thus results in shrinking the lattice parameters. Mizuguchi et al. make similar arguments on the moisture-induced superconductivity in $FeTe_{0.8}S_{0.2}$ [13]. Another likely scenario is the substitution of the $O^{2-}$ ions by the $(OH)^-$ ions. Their ionic radii are 1.40 Å and 1.37 Å, respectively. Accordingly, substitution of $O^{2-}$ by $(OH)^-$ should result in contracting the lattice parameters. Further studies are needed to understand the role of H atoms in the *Ln*1111(H) system.

In summary, we have discovered that $T_c$ of *Ln*1111 (*Ln* = La, Ce and Pr) increases by incorporating hydrogen. The present results not only propose another route in increasing $T_c$ of the Fe-based superconductors but also reinforce the intimate relationship between $T_c$ and the lattice parameters in the Fe-based superconductors.

**Acknowledgements:**

We thank Dr. I. Yamada, Dr. Y. Takano, Prof. Y. Kitaoka and Prof. M. Takano for enlightening discussions. This work was supported by Grant-in-Aid for Specially promoted Research (20001004) from The Ministry of Education, Culture, Sports, Science and Technology (MEXT), Mitsubishi Foundation and JST, Transformative Research-Project on Iron Pnictides (TRIP).

**Figure captions:**

Fig. 1 Sample cell assembly used in HP synthesis technique: (a) method-I and (b) method-II. (c) XRD patterns of *Ln*1111(H) for *Ln* = La (method-I and II), Ce and Pr (method-II).

Fig. 2 $^{1}$H-NMR spectrum of (a) La1111(H) ($H$ = 3.916 T, $T$ = 100 K) and (b) $Ag_6O_8Ag(HF_2)$ ($H$ = 3.184 T, $T$ = 110 K) [9].

Fig. 3 Temperature dependent susceptibility for (a) La1111(H) (method-I), (b) La1111(H) (method-II), (c) Ce1111(H) (method-II) and (d) Pr1111(H) (method-II). Data of the H-free O-deficient samples $LnFeAsO_{1-y}$ (*Ln*1111) are also shown in the figures as open circles. The base lines (normal state susceptibility) are shifted for comparison. $T_c$ are indicated by arrows.

Fig. 4 Temperature dependent resistivity for *Ln*1111(H) (*Ln*=La, Ce and Pr, method-II). The inset shows the temperature dependence of resistivity near $T_c$.

Fig. 5 *a*- and *c*-axes lattice parameters and $T_c$ ($\chi$-onset) of the *Ln*1111(H) samples together with those of H-free O-deficient *Ln*1111 for comparison. The inset shows that relationship between the *a*-axis lattice parameter and $T_c$ ($\chi$-onset) for the highest $T_c$ H-free O-deficient *Ln*1111 and *Ln*1111(H). Dashed lines are guides to the eye.

Table I Lattice parameters and $T_c$ of *Ln*1111(H) (*Ln* = La, Ce and Pr).

| Samples | *a* (Å) | *c* (Å) | $T_c$ (K) ($\rho$-onset) | $T_c$ (K) ($\rho$-zero) | $T_c$ (K) ($\chi$-onset) | Methods |
|---|---|---|---|---|---|---|
| La1111(H) | 3.9876(2) | 8.6687(9) | - | - | 35.0 | I |
| La1111(H) | 3.9938(2) | 8.6898(9) | 38.5 | 34.0 | 35.0 | II |
| Ce1111(H) | 3.9740(2) | 8.5953(8) | 48.0 | 44.6 | 45.8 | II |
| Pr1111(H) | 3.9522(2) | 8.5608(8) | 51.9 | 49.0 | 49.7 | II |

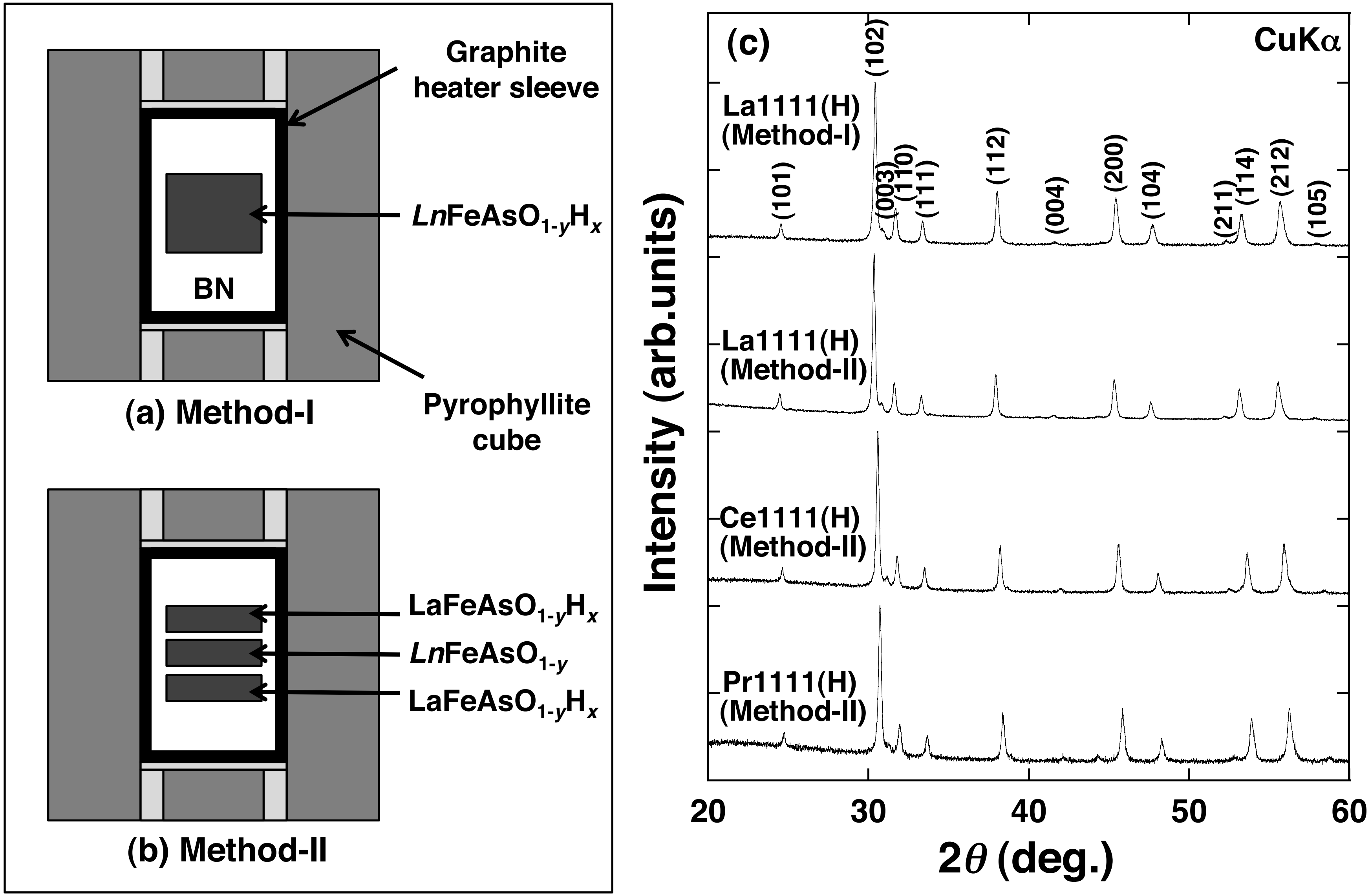

Fig.1 Miyazawa et al.

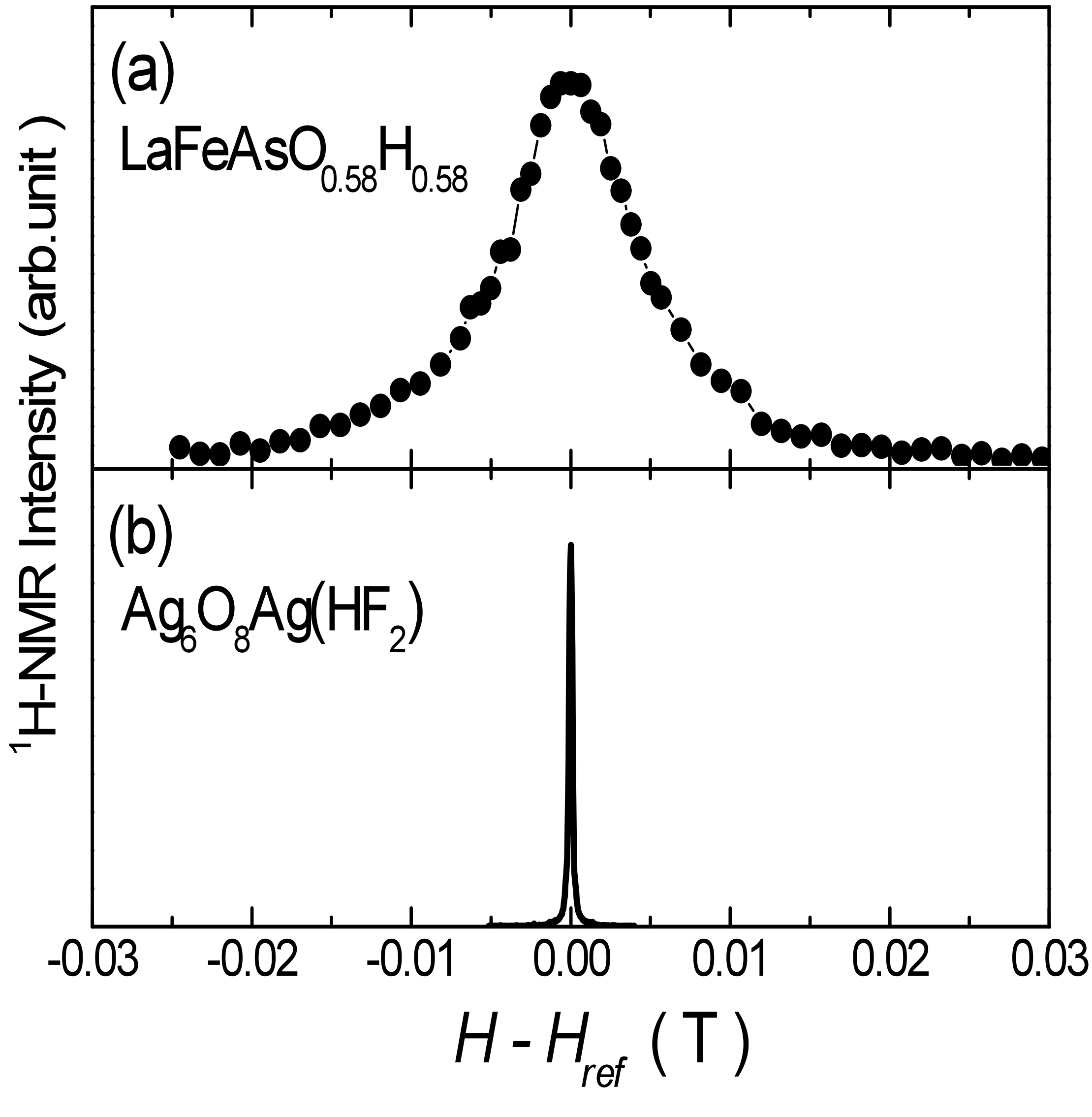

Fig. 2 Miyazawa et al.

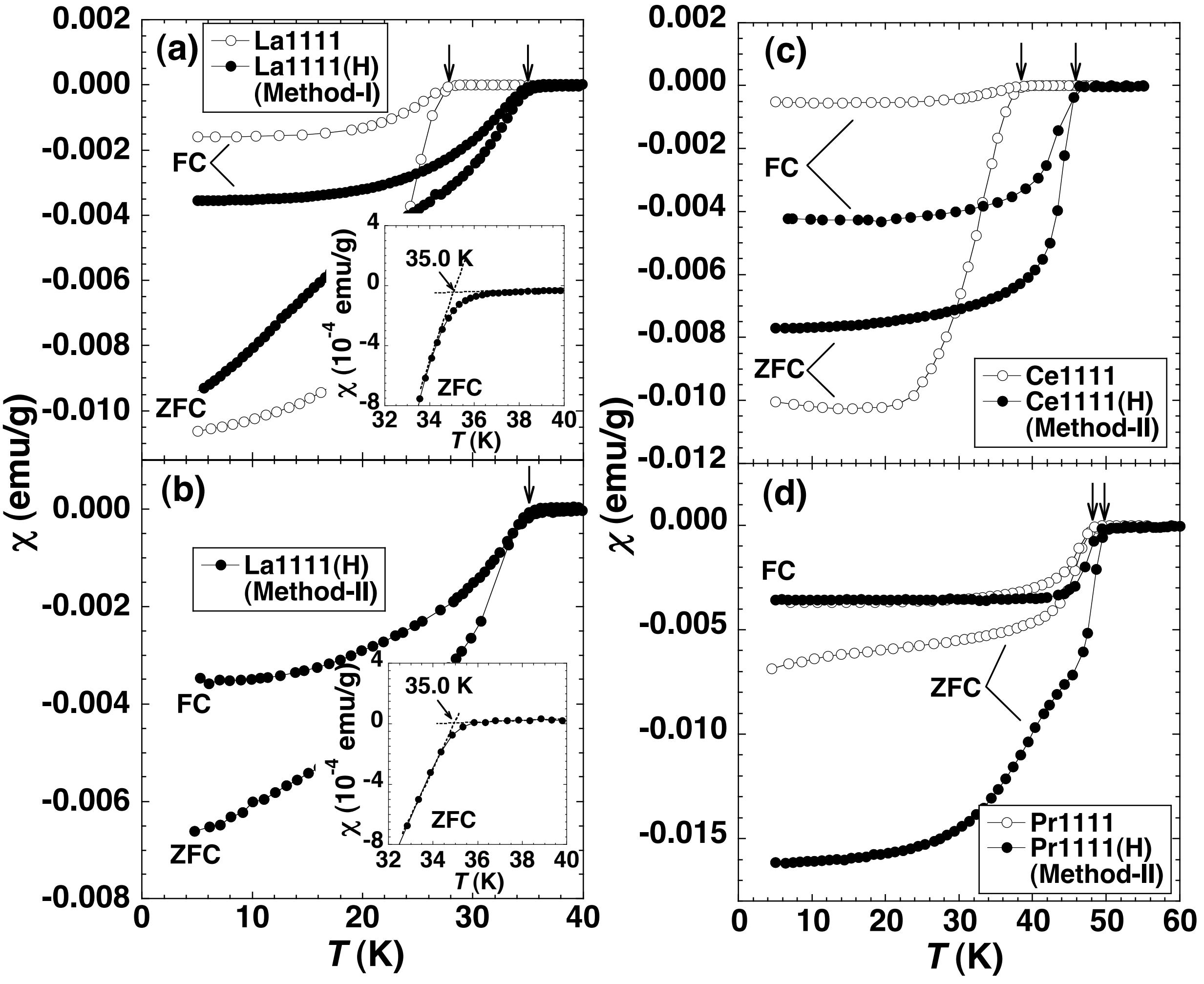

Figure 3 Miyazawa et al.

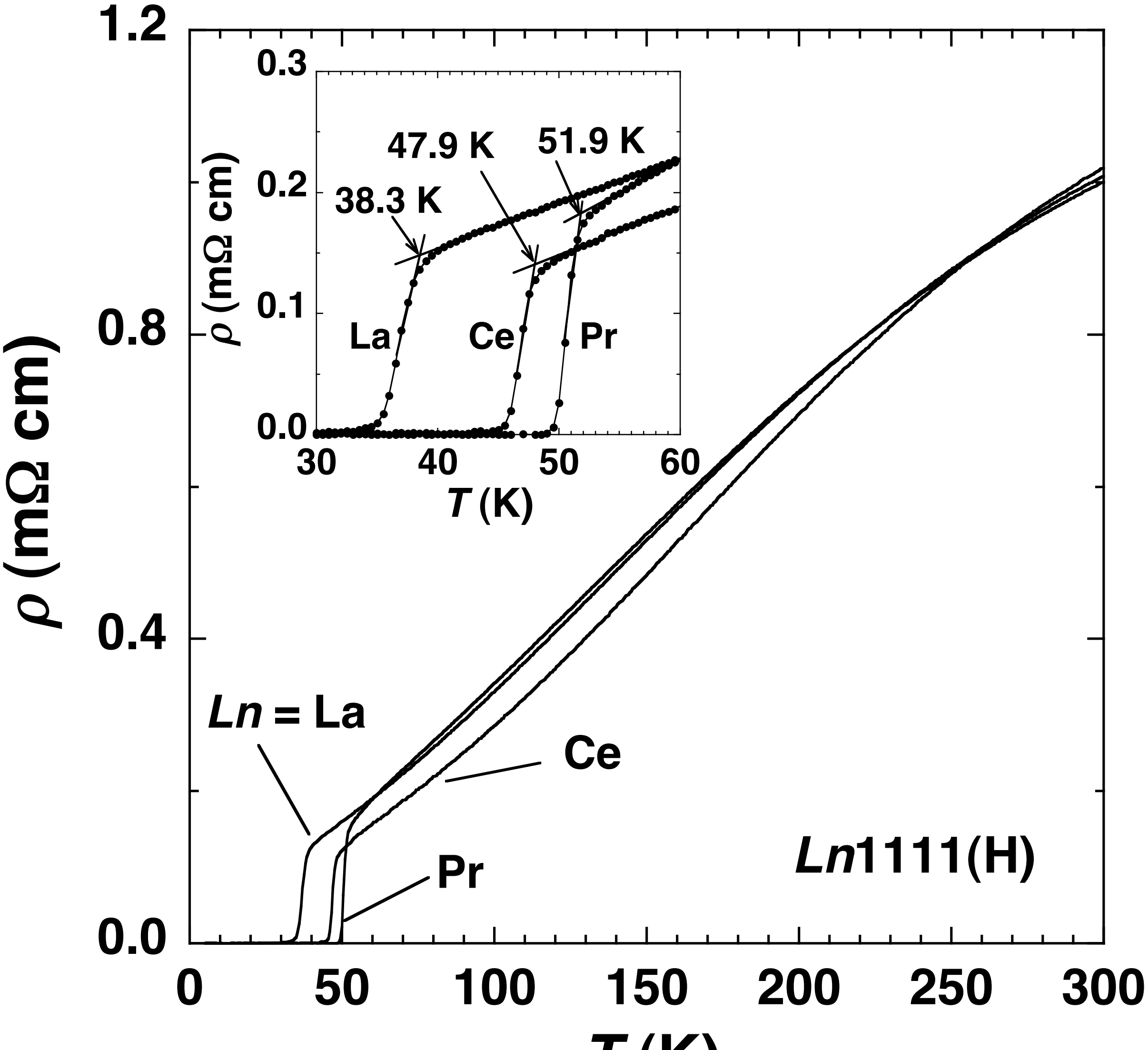

Figure 4 Miyazawa et al.

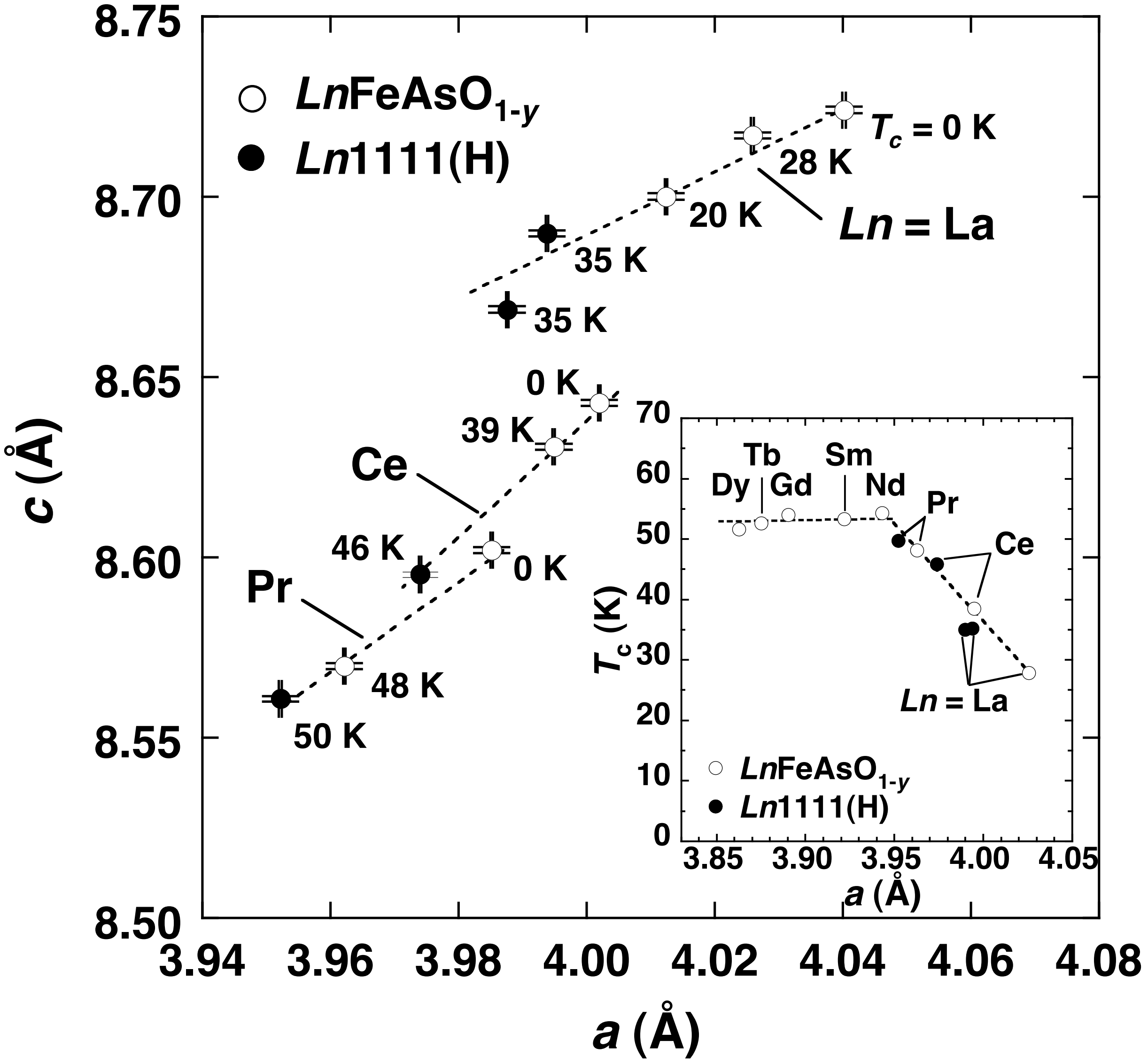

Fig. 5 Miyazawa et al.